\documentclass[12pt]{article}

\usepackage[T1]{fontenc}
\usepackage{amssymb}
\usepackage{graphicx}
\usepackage{amsmath}
\usepackage{tensor}
\usepackage{hyperref}
\usepackage{epstopdf}
\usepackage{extarrows}
\usepackage{xcolor}
\usepackage{indentfirst}
\newcommand{\be}{\begin{equation}}
\newcommand{\ee}{\end{equation}}
\newcommand{\ba}{\begin{aligned}}
\newcommand{\ea}{\end{aligned}}
\newcommand{\been}{\begin{eqnarray*}}
\newcommand{\een}{\end{eqnarray*}}
\newcommand{\beg}{\begin{equation}\begin{aligned}}
\newcommand{\eng}{\end{aligned}\end{equation}}

\def\be {\begin{equation}}
\def\ee {\end{equation}}

\def\be{\begin{equation}}
\def\ee{\end{equation}}
\def\bea{\begin{eqnarray}}
\def\eea{\end{eqnarray}}

\def\l{\lambda}

\def\>{\rangle} 
\def\<{\langle} 
\def\vev#1{\langle #1\rangle}

\begin{document}

\begin{titlepage}
\vspace{0.5cm}
\begin{center}
{\Large \bf Note on higher-point correlation functions of the $T\bar{T}$ or $J\bar{T}$ deformed CFTs}

\lineskip .75em
\vskip 2.5cm
{\large Song He$^{a,b,}$\footnote{hesong@jlu.edu.cn}}
\vskip 2.5em
 {\normalsize\it $^{a}$Center for Theoretical Physics, College of Physics, Jilin University, Changchun 130012, People's Republic of China\\
 $^{b}$Max Planck Institute for Gravitational Physics (Albert Einstein Institute),\\
Am M\"uhlenberg 1, 14476 Golm, Germany}\\
\vskip 3.0em
\end{center}
\begin{abstract}
      We investigate generic n-point correlation functions of conformal field theories (CFTs), with $T\bar{T}$ and $J\bar{T}$ deformations, in terms of the perturbative CFT approach. We systematically obtain the first order correction to the generic correlation functions of CFTs with $T\bar{T}$ or $J\bar{T}$  deformation. We compute the out of time ordered correlation function (OTOC) in the Ising model with $T\bar{T}$ or $J\bar{T}$ deformation, which  confirms that these deformations do not change the integrable property up to the first order level.
\end{abstract}
\end{titlepage}

\tableofcontents
\section{Introduction}
The class of exactly solvable deformations of 2D QFTs with rotational and translational symmetries, called $T\bar{T}$ deformation \cite{Zamolodchikov:2004ce,Smirnov:2016lqw,Cavaglia:2016oda}, attracts a large amount of attention in research. $T\bar{T}$ deformation has many intriguing properties,
but this kind of irrelevant deformation is usually hard to handle. One remarkable property is its integrability \cite{Smirnov:2016lqw,Cavaglia:2016oda,LeFloch:2019wlf,Jorjadze:2020ili,Datta:2018thy,Aharony:2018bad,Rosenhaus:2019utc}. If the un-deformed theory is integrable, there exists a set of infinite commuting conserved charges or KdV charges \cite{LeFloch:2019wlf} in the deformed theory. These deformations also preserve the integrable properties of the un-deformed theory.

For $T \bar{T}$ deformation, it was proposed that the $T \bar{T}$ deformed CFT corresponded to cutoff $\operatorname{AdS}_{3}$ at a finite radius with the Dirichlet boundary condition \cite{McGough:2016lol,Giribet:2017imm,Kraus:2018xrn}. There are some non-trivial checks for this proposal. The $J\bar{T}$ deformation also has a holographic interpretation \cite{Bzowski:2018pcy,Chakraborty:2018vja,Nakayama:2019mvq,Roychowdhury:2020zer}.
Moreover, these deformations are also related to string theory \cite{Chakraborty:2018vja,Giveon:2017nie,Dei:2018mfl, Araujo:2018rho,Frolov:2019nrr,Callebaut:2019omt,Tolley:2019nmm,Chakraborty:2019mdf,Apolo:2018qpq,Sfondrini:2019smd,Apolo:2019yfj,Apolo:2019zai}.
 The gravitational aspects associated with $T\bar{T}$ deformation have been studied by \cite{Dubovsky:2017cnj,Cardy:2018sdv,Dubovsky:2018bmo,Ishii:2019uwk,Okumura:2020dzb}. The gravitational perturbations can be regarded as the $T\bar{T}$ deformations of 2D QFT.
These deformed theories can be well controlled by the fact that many quantities in the deformed theory, such as: the S-matrix, energy spectrum, Wilson loop, correlation functions, entanglement entropy, can be computed analytically \cite{Smirnov:2016lqw,Rosenhaus:2019utc,Cardy:2019qao,Donnelly:2018bef,Chen:2018eqk,Sun:2019ijq,Jeong:2019ylz,Chakraborty:2018aji,Hirano:2020ppu}, in particular when the un-deformed theory is a CFT. These deformations also attract attention from both field theory \cite{Guica:2017lia,Bonelli:2018kik,Cardy:2018jho,Santilli:2018xux,Conti:2019dxg,Baggio:2018rpv,Chang:2018dge,Jiang:2019trm,Jiang:2019hux,Chang:2019kiu,Coleman:2019dvf,Ebert:2020tuy}, and the holographic perspective \cite{Giveon:2017myj,Asrat:2017tzd,He:2019glx,Lewkowycz:2019xse,Chen:2019mis,Geng:2019yxo,Geng:2019ruz,Guica:2019nzm,Li:2020pwa,Ouyang:2020rpq,Santilli:2020qvd,Asrat:2020jsh}.

 The correlation functions are fundamental observables in QFTs, therefore, it is important to study them in their own right. The correlation functions have many significant applications, e.g., quantum chaos and quantum entanglement. In particular, four-point functions are associated with the out-of-time-order correlator (OTOC), which can be applied to diagnose chaotic behavior in field theory with/without the deformations \cite{Roberts:2014ifa, Shenker:2014cwa, Maldacena:2015waa,He:2019vzf}. To measure quantum entanglement, the computation of entanglement (or R\'{e}nyi) entropies involves correlation functions \cite{Calabrese:2004eu}. In particular, one can apply the higher point correlation function to calculate the R\'{e}nyi entanglement entropy of the local excited states in 2D conformal field theory for various situations \cite{He:2014mwa,He:2017lrg}.  The $T\bar{T}$ deformed partition function, namely the zero-point correlation function, on a torus can be computed and shown to be a modular invariant \cite{Datta:2018thy,Aharony:2018bad}. Furthermore, the partition function, {with chemical potentials for KdV charges turning on}, was also obtained by \cite{Asrat:2020jsh}. The correlation functions with $T\bar{T}$ and $J\bar{T}$ deformations in the deep UV theory were investigated in a non-perturbative way by \cite{Cardy:2019qao}.

In the present work, we are interested in studying the higher point correlation functions in the $T\bar{T}$ and  $J\bar{T}$ deformed CFTs. Here, we focus on the deformation region near the un-deformed CFTs, where the CFT Ward identity holds, and, in the current setup, the renormalization group flow effect of the operator with irrelevant deformation can be neglected. The total Lagrangian is expanded near the critical point for a small coupling constant $\l$
\be
\mathcal{L}=\mathcal{L}_{CFT}-\l \int d^2z O(z,\bar z).
\ee
The first order correction to the deformed correlation function takes the following form
\be\label{eq1}
\l\int_{C} d^2z\vev{O(z,\bar z)\phi_1(z_1)...\phi_n(z_n)},
\ee
where $O(z,\bar z)$ can be $T\bar{T}$ or $J\bar{T}$, and the expectation value in the integrand is calculated in the underformed CFTs using the Ward identity.
In the perturbative CFT approach, the deformed two-point functions and three-point functions were considered in \cite{Kraus:2018xrn,Guica:2019vnb} up to the first order of coupling constant. Subsequently, we have previously considered the four-point functions on the plane \cite{He:2019vzf} and on the torus \cite{He:2020udl}. Also, we have generalized this study to the case with supersymmetric extension \cite{He:2019ahx}. More recently, the $T\bar{T}$ flow effect has been taken into account in the computation of the partition function of CFTs on tori in the Lagrangian formalism up to the second order deformation \cite{He:2020cxp}. In the present work, we would like to follow the same approach to obtain the generic n-point correlation function of the $T\bar{T}$ and $J\bar{T}$ of the deformed field theories. Since the n-point correlation functions depend on the $2n-3$ holomorphic and anti-holomorphic cross ratios, one can also apply the Ward identity to obtain the first order correction to the deformed correlation function, which is quite complicated. Since the OTOC can diagnose quantum chaos, in particular, the late time behavior of OTOC gives strong evidence to confirm whether the system is maximum chaos or integrable. {The paper \cite{He:2019vzf} has extracted the large time behavior in large c CFT with a large c expansion, to show that the $T\bar{T}$ and $J\bar{T}$ deformations do not change the maximal quantum chaos properties of the un-deformed large central charge CFT that has a holographic dual description. It is natural to ask whether these deformations preserve the integrability properties of the un-deformed integrable CFTs. In AdS/CFT, there are many promising and practical proposals for investigating the quantum chaos, by analyzing the spectral form factor and OTOC to show early time and late time chaos. By empiricism, one can calculate these quantities in various quantum field theories to extract the universal behavior of quantum chaos. Since the Ising model is the simplest integrable CFT with a finite central charge, one can study the OTOC in the Ising model to confirm two things. Firstly, that the deformations preserve the integrability properties of the original integrable CFTs from a quantum information point of view. Secondly, to check whether OTOC is a good phenomenological quantity to capture the quantum chaos or integrability. In this paper, we take the 2D Ising model as a simple example to check whether deformations preserve the integrability property of the un-deformed CFT with full central charge corrections.}

The structure of this paper is as follows. In Section 2, we review the generic n-point correlation function of the CFTs, and we apply  the Ward identity associated with $T$, $\bar{T}$, and $J$ to study the first order correction of the $T\bar{T}$ or $J\bar{T}$ deformed n-point correlation function. In Section 3, we apply the deformed correlation function to study the OTOC in the Ising model to show that the $T\bar{T}$ and $J\bar{T}$ deformations do not change the integrability property of the un-deformed Ising model. Conclusions and discussions are given in the final section. In the Appendix, we list some relevant techniques and notation that are useful in our analysis.

\section{n-point correlation functions in the deformed CFTs}
In this section, we review the generic structure of the n-point correlation function in 2D CFTs. We apply these structures to construct the first order correction to the $T\bar{T}$ or $J\bar{T}$ deformed correlation function.
Using the constraints of global conformal invariance, the $n$-point function \cite{Ginsparg:1988ui} in CFT can be written in the following form
\be\label{eqn}
\vev{O_1(z_1,\bar{z}_1)...O_n(z_n,\bar{z}_n)}=f(\eta_i,\bar{\eta}_i)\prod_{i<j}^n z_{ij}^{-a_{ij}}\bar{z}_{ij}^{-\bar{a}_{ij}},
\ee
where $\eta_i$ are the $n-3$ cross ratios and \be f(\eta_1,...\eta_i,...\eta_{n-2},\bar{\eta}_1,...\bar{\eta}_i,...,\bar{\eta}_{n-2})\nonumber\ee is abbreviated as $f(\eta_i,\bar{\eta}_i)$.
$a_{ij}$ and the individual conformal dimension $h_i$ of each operator eq.(\ref{eqn}) satisfies the $n$ equations
\be
2h_i=\sum_{i<j} a_{ij}+\sum_{i>j} a_{ji}.
\ee
One special solution is
\be
a_{ij}=\frac{2}{n-2}\left(\frac{h_t}{n-1}-h_i-h_j\right),~~h_t=\sum_i^n h_i.
\ee
By global conformal transformation
\be
z\to \frac{(z-z_1)(z_{n-1}-z_n)}{(z-z_n)(z_{n-1}-z_1)},
\ee
where there are two (holomorphic) independent cross ratios\footnote{
	The number of independent cross ratios in $D$-dimensional spacetime is
	\be
	\min(C_n^2-n,nD-\frac{(D+2)(D+1)}{2}).
	\ee
	Take $n=5,D=2$, for example. The two independent cross ratios can be chosen as $A_2,A_3$
	\be
	A_3=\frac{z_{12}z_{45}}{z_{25}z_{14}}\to A'_3=\frac{z_{15}z_{42}}{z_{25}z_{14}},~~A_2=\frac{z_{13}z_{45}}{z_{35}z_{14}}\to A'_2=\frac{z_{15}z_{43}}{z_{35}z_{14}},
	\ee
	then \be
	A_3/A_2\to A_4=\frac{z_{12}z_{35}}{z_{25}z_{13}}\to A'_4=\frac{z_{15}z_{23}}{z_{25}z_{13}},
	\ee
	\be
	A'_3/A'_4\to A_5,~~A'_3/A'_3\to A_1.
	\ee} for $n=5$ and 3 for $n=6$ in 2D

\be
\eta_i=\frac{(z_i-z_1)(z_{n-1}-z_n)}{(z_i-z_n)(z_{n-1}-z_1)},~~2\leq i\leq n-2,
\ee
We are interested in the cases with equal conformal dimension $h_i$, so
\be
\vev{O_1(z_1,\bar{z}_1)...O_n(z_n,\bar{z}_n)}=f(\eta_i,\bar{\eta}_i)\prod_{i<j}^n z_{ij}^{-a_{ij}}\bar{z}_{ij}^{-\bar{a}_{ij}},~~a_{ij}=-\frac{2h}{n-1},
\ee
with all $h_{t}=nh$ and $a_{ij}\equiv a$.
\subsection{n-point correlation function in $T\bar{T}$ deformed CFTs}
In this subsection, we construct the first order correction to the n-point correlation function in the $T\bar{T}$ deformed CFTs
\be
\l\int_{C} d^2z\vev{T\bar{T}(z,\bar z)\phi_1(z_1)...\phi_n(z_n)}.
\ee
For simplicity, we define the following symbols
\begin{align}
O_{L}&=\prod_{i<j}z_{ij}^{-a}\qquad\qquad\qquad
O_{R}=\prod_{i<j}\bar{z}_{ij}^{-{a}}\\
F&=\sum_{i=1}^{n}\frac{h}{(z-z_{i})^{2}}\qquad\qquad
\bar{F}=\sum_{i=1}^{n}\frac{\bar{h}}{(\bar{z}-\bar{z}_{i})^{2}}\\
G&=\sum_{i=1}^{n}\frac{\partial_{z_{i}}}{z-z_{i}}\qquad\qquad\quad
\bar{G}=\sum_{i=1}^{n}\frac{\partial_{\bar{z}_{i}}}{\bar{z}-\bar{z}_{i}}\\
T&=F+G\qquad\qquad\qquad\quad
\bar{T}=\bar{F}+\bar{G}
\end{align}
Using the conformal Wald identity, the single energy momentum tensor $T$ acting on the generic n-point correlation function can be written as
\begin{align}\label{singleTaction}
&\langle TO_{n}\rangle
=FO_{n}
+(GO_{L})f(\eta,\bar{\eta})O_{R}
+(GO_{R})f(\eta,\bar{\eta})O_{L}
+(Gf(\eta,\bar{\eta})){O_{L}}O_{R}.
\end{align}
Since
\begin{align}
&\frac{1}{z-z_{i}}\frac{\partial}{\partial z_{i}}O_{L}=\frac{a}{z-z_{i}}\sum_{j\neq i}\frac{1}{z_{ji}}O_{L},
\end{align}
then
\begin{align}
GO_{L}=\left(\sum_{i=1}^{n}\sum_{j\neq i}\frac{a}{z-z_{i}}\frac{1}{z_{ji}}\right)O_{L}=pO_{L},
\end{align}
with
\begin{align}
p=\sum_{i=1}^{n}\sum_{j\neq i}\frac{a}{z-z_{i}}\frac{1}{z_{ji}}.
\end{align}
The main factor in the third term of eq.(\ref{singleTaction}) is
\begin{align}
&\sum_{i=1}^{n}\frac{1}{z-z_{i}}\frac{\partial}{\partial z_{i}}O_{R}
=\frac{2\pi{a}}{z-z_{i}}\sum_{k\neq1}\bar{z}_{ik} O_{R} \delta^{(2)}(z_{ik}).
\end{align}
Since $\forall i,k$, always $\exists s,t, \text{such that  } s=k,t=i$
\begin{align}
\frac{1}{z-z_{i}}\bar{z}_{ik}\delta^{(2)}(z_{ik})+\frac{1}{z-z_{s}}\bar{z}_{st}\delta^{(2)}(z_{st})=0,
\end{align}
then
\begin{align}
&\sum_{i=1}^{n}\frac{1}{z-z_{i}}\frac{\partial}{\partial z_{i}}O_{R}=0.
\end{align}
One can check the main factor in the third term $Gf(\eta,\bar{\eta})$ of  eq.(\ref{singleTaction}) is
\begin{align}\label{eq33}
&\sum_{i=1}^{n}\frac{{\partial_{z_i}}}{z-z_{i}}\sum_{j=2}^{n-2}f(\eta_{j},\bar{\eta_{j}})={q\over f},
\end{align}
where \be
\ba
q=\sum_{j=2}^{n-2}\eta_{j}\frac{\partial f(\eta_{j},\bar{\eta}_{j})}{\partial\eta_{j}}\tilde{q}.
\ea
\ee
For generic $n$, one can obtain the following result
\be
\ba
F+p=\frac{h}{n-1}\sum_{i=1}^{n}\sum_{j>i}\frac{z_{ij}^{2}}{(z-z_{i})^2(z-z_{j})^2}.
\ea
\ee
Finally, the total contribution to the first order deformation of the correlation function with a single operator insertion $T$ is
\begin{align}
\langle TO_{n}\rangle=(F+p+{q\over f})\langle O_{n}\rangle.
\end{align}
The explicit form is the following
\be
\ba
\langle TO_{n}\rangle&=(F+p)fO_{L}O_{R}+\sum_{j=2}^{n-2}\eta_{j}\frac{\partial f(\eta_{j},\bar{\eta}_{j})}{\partial\eta_{j}}\tilde{q}O_{L}O_{R},
\ea
\ee
with
\be
\tilde{q}=\frac{1}{z-z_{1}}\frac{z_{j,n-1}}{z_{n-1,1}z_{j1}}+\frac{1}{z-z_{j}}\frac{z_{1n}}{z_{nj}z_{1j}}+\frac{1}{z-z_{n-1}}\frac{z_{n1}}{z_{1,n-1}z_{n,n-1}}+\frac{1}{z-z_{n}}\frac{z_{n-1,j}}{z_{jn}z_{n-1,n}}.
\ee
Similarly, one can also find that
\be\label{singleTinsertion}
\langle\bar{T}O_{n}\rangle=(\bar{F}+\bar{p})fO_{L}O_{R}+\sum_{j=2}^{n-2}\bar{\eta}_{j}\frac{\partial f(\eta_{j},\bar{\eta}_{j})}{\partial\bar{\eta}_{j}}\bar{\tilde{q}}O_{L}O_{R}.
\ee
As a consistency check, the correlation function with one single operator insertion $T$ in the $n=2,3, 4$ case is the same as the one presented in \cite{He:2019vzf}. In particular, the two- and three-point correlation functions have nothing to do with the final term in eq.(\ref{singleTinsertion}).
For $n=2$, we have
$$
\begin{aligned}
F+p &=\frac{h z_{12}^{2}}{\left(z-z_{1}\right)^{2}\left(z-z_{2}\right)^{2}},
\end{aligned}
$$
which is the same as the two-point correlation function inserted with a single $T$ in \cite{He:2019vzf}.
For $n=3,$ we have
$$
\begin{aligned}
F+p=\frac{h_{12}^{2}}{2}\left(\frac{z_{13}^{2}}{\left(z-z_{1}\right)^{2}\left(z-z_{2}\right)^{2}}+\frac{z_{23}^{2}}{\left(z-z_{1}\right)^{2}\left(z-z_{3}\right)^{2}}+\frac{1}{\left(z-z_{2}\right)^{2}\left(z-z_{3}\right)^{2}}\right)
\end{aligned}
$$
For $n=4$, we use a slightly different notation of the four-point function between the current study and  \cite{He:2019vzf}
\be
\begin{aligned}\label{notation}
\left\langle O_{4}\right\rangle_{\text{In our paper}} &=f(\eta, \bar{\eta}) \prod_{i<j} z_{i j}^{-a} \bar{z}_{i j}^{-\bar{a}} \\
\left\langle O_{4}\right\rangle_{\text{Notation in \cite{He:2019vzf}}} &=\tilde{f}(\eta, \bar{\eta}) \frac{1}{z_{13}^{2 h} z_{24}^{2 h} \bar{z}_{13}^{2 \bar{h}} \bar{z}_{24}^{2 \bar{h}}}
\end{aligned}
\ee
where the $\tilde{f}(\eta, \bar{\eta})=\eta^{-{2h \over 3}}\bar{\eta}^{-{2h \over 3}}(1-\eta)^{-{2h\over 3}}(1-\bar{\eta})^{-{2h\over 3}}{f}(\eta, \bar{\eta})$. Finally, one can reproduce the four-point correlation function with a single inserted $T$ in \cite{He:2019vzf}.

One can obtain the full expression of {$\langle T\bar{T}O_{n}\rangle$}
\be
\ba
\langle T\bar{T}O_{n}\rangle&=(\bar{F}+\bar{G})\Big((F+p)fO_{L}O_{R}+\sum_{j=2}^{n-2}\eta_{j}\frac{\partial f(\eta_{j},\bar{\eta}_{j})}{\partial\eta_{j}}\tilde{q}O_{L}O_{R}\Big)\\
&=(F+p)\Big((\bar{F}+\bar{p})\langle O_{n}\rangle +\sum_{j=2}^{n-2}\frac{\bar{\eta}_{j}}{f(\eta_{j},\bar{\eta}_{j})}\frac{\partial f(\eta_{j},\bar{\eta}_{j})}{\partial\bar{\eta}_{j}}\bar{\tilde{q}}\langle O_{n}\rangle\Big)\\&+\bar{F}\sum_{j=2}^{n-2}\eta_{j}\frac{1}{f(\eta_{j},\bar{\eta}_{j})}\frac{\partial f(\eta_{j},\bar{\eta}_{j})}{\partial\eta_{j}}\tilde{q}\langle O_{n}\rangle\\
&=(F+p)(\bar{F}+\bar{p})\langle O_{n}\rangle+(F+p)\sum_{j=2}^{n-2}\frac{\bar{\eta}_{j}}{f(\eta_{j},\bar{\eta}_{j})}\frac{\partial f(\eta_{j},\bar{\eta}_{j})}{\partial\bar{\eta}_{j}}\bar{\tilde{q}}\langle O_{n}\rangle\\
&+(\bar{F}+\bar{p})\sum_{j=2}^{n-2}\frac{\eta_{j}}{f(\eta_{j},\bar{\eta}_{j})}\frac{\partial f(\eta_{j},\bar{\eta}_{j})}{\partial\eta_{j}}\tilde{q}O_{n}+\bar{G}(\sum_{j=2}^{n-2}\eta_{j}\frac{\partial f(\eta_{j},\bar{\eta}_{j})}{\partial\eta_{j}}\tilde{q})O_{L} O_{R}\\&+\Big(\bar{G}(F+p)\Big)\langle O_{n}\rangle,
\ea
\ee
where we have used $\bar{G}O_{L}=0,\bar{G}O_{R}=\bar{p}O_R$ and $f(\eta,\bar{\eta})O_{L}O_{R}=\langle O_n \rangle$.

Since
\be
\label{eq:q-tilde}
\sum_{i=1}^n\frac{1}{z-z_{i}}\frac{\partial\eta_{j}}{\partial z_{i}}=\eta_{j}\tilde{q},
\ee
we rewrite the first order deformation of the n-point correlation function as follows
\be\label{TTdeformation}
\ba
\langle\bar{T}TO_{n}\rangle&=\int d^{2}z\Big((F+p)(\bar{F}+\bar{p})+(F+p)\sum_{j=2}^{n-2}\frac{\bar{\eta}_{j}}{f(\eta_{j},\bar{\eta}_{j})}\frac{\partial f(\eta_{j},\bar{\eta}_{j})}{\partial\bar{\eta}_{j}}\bar{\tilde{q}}\\&+(\bar{F}+\bar{p})\sum_{j=2}^{n-2}\frac{\eta_{j}}{f(\eta_{j},\bar{\eta}_{j})}\frac{\partial f(\eta_{j},\bar{\eta}_{j})}{\partial\eta_{j}}\tilde{q}+\frac{1}{f(\eta_{j},\bar{\eta}_{j})}\sum_{j,{\tilde{j}}=2}^{n-2}\frac{\partial^{2}f(\eta_{j},\bar{\eta}_{j})}{\partial\eta_{j}\partial\bar{\eta}_{{\tilde{j}}}}\eta_{j}\tilde{q}\bar{\eta}_{{\tilde{j}}}\bar{\tilde{q}}\Big)\langle O_{n}\rangle.
\ea
\ee
For later convenience, we can define the following conventions
\be\ba
&\l\int_{C} d^2z\vev{T\bar{T}(z,\bar z)\phi_1(z_1)...\phi_n(z_n)}\\&=\Big(G^{T\bar T}_1+G^{T\bar T}_2+G^{T\bar T}_3+G^{T\bar T}_4\Big)\langle O_{n}\rangle,\\
G^{T\bar T}_1&=\int d^{2}z(F+p)(\bar{F}+\bar{p}),\\
G^{T\bar T}_2&=\int d^{2}z(F+p)\sum_{\tilde{j}=2}^{n-1}\frac{\bar{\eta}_{\tilde{j}}}{f}\frac{\partial f}{\partial\bar{\eta}_{\tilde{j}}}\bar{\tilde{q}},\\
G^{T\bar T}_3&=(\bar{F}+\bar{p})\sum_{j=2}^{n-2}\frac{\eta_{j}}{f(\eta_{j},\bar{\eta}_{j})}\frac{\partial f(\eta_{j},\bar{\eta}_{j})}{\partial\eta_{j}}\tilde{q},\\
G^{T\bar T}_4&=\frac{1}{f(\eta_{j},\bar{\eta}_{j})}\sum_{j,{\tilde{j}}=2}^{n-2}\frac{\partial^{2}f(\eta_{j},\bar{\eta}_{j})}{\partial\eta_{j}\partial\bar{\eta}_{{\tilde{j}}}}\eta_{j}\tilde{q}\bar{\eta}_{{\tilde{j}}}\bar{\tilde{q}}.\\
\ea\ee
The first term of eq.(\ref{TTdeformation}) is as follows
\be\label{first-term}
\ba
G^{T\bar T}_1=&\int d^{2}z(F+p)(\bar{F}+\bar{p})\\=&\Big(\frac{h}{n-1}\Big)^2\sum_{i=1}^{n}\sum_{j>i}\sum_{\tilde{i}=1}^{n}\sum_{\tilde{j}>\tilde{i}}\int d^{2}z\frac{z_{ij}^{2}}{(z-z_{i})^2(z-z_{j})^2}\frac{\bar{z}_{\tilde{i}\tilde{j}}^{2}}{(\bar{z}-\bar{z}_{\tilde{i}})^2(\bar{z}-\bar{z}_{\tilde{j}})^2}
\ea
\ee
and this integration can be simplified as eq.(\ref{firstTT})in the appendix A.1.

The second term of eq.(\ref{TTdeformation}) is can be simplified as following\footnote{One can refer to the details in appendix A.2.}
\be
\ba\label{secondTT}
G^{T\bar T}_2 =&\int d^{2}z(F+p)\sum_{\tilde{j}=2}^{n-2}\frac{\bar{\eta}_{\tilde{j}}}{f}\frac{\partial f}{\partial\bar{\eta}_{\tilde{j}}}\bar{\tilde{q}}\\
=&\frac{h}{n-1}\sum_{\tilde{j}=2}^{n-2}\frac{\bar{\eta}_{\tilde{j}}}{f}\frac{\partial f}{\partial\bar{\eta}_{\tilde{j}}}\Big(\sum_{i=1,i\neq1}^{n}\sum_{j>i}{z_{ij}^{2}}{\cal I}_{221}(z_{i},z_{j},\bar{z}_{1})\frac{\bar{z}_{\tilde{j},n-1}}{\bar{z}_{n-1,1}\bar{z}_{\tilde{j}1}}\\
+&\sum_{i=1,i\neq\tilde{j}}^{n}\sum_{j>i,j\neq\tilde{j}}{z_{ij}^{2}}{\cal I}_{221}(z_{i},z_{j},\bar{z}_{\tilde{j}})\frac{\bar{z}_{1,n}}{\bar{z}_{n,\tilde{j}}\bar{z}_{1\tilde{j}}}
+\sum_{i=1,i\neq\tilde{j}}^{n}\sum_{j=\tilde{j}>i}{z_{ij}^{2}}{\cal I}_{221}(z_{i},z_{\tilde{j}},\bar{z}_{\tilde{j}})\frac{\bar{z}_{1n}}{\bar{z}_{n\tilde{j}}\bar{z}_{1\tilde{j}}}\\
+&\sum_{i=1,i\neq n-1}^{n}\sum_{j>i,j\neq n-1}{z_{ij}^{2}}{\cal I}_{221}(z_{i},z_{j},\bar{z}_{n-1})\frac{\bar{z}_{n}-\bar{z}_{1}}{\bar{z}_{1,n-1}\bar{z}_{n,n-1}}\\+&\sum_{i=1,i\neq n-1}^{n}\sum_{j=n-1>i}{z_{i,n-1}^{2}}{\cal I}_{221}(z_{i},z_{n-1},\bar{z}_{n-1})\frac{\bar{z}_{n}-\bar{z}_{1}}{\bar{z}_{1,n-1}\bar{z}_{n,n-1}}\\+&\sum_{i=1}^{n}\sum_{j>i,j\neq n}{z_{ij}^{2}}{\cal I}_{221}(z_{i},z_{j},\bar{z}_{n})\frac{\bar{z}_{n-1,\tilde{j}}}{\bar{z}_{\tilde{j}n}\bar{z}_{n-1,n}}
+\sum_{i=1}^{n}\sum_{j=n>i}{z_{in}^{2}}{\cal I}_{221}(z_{i},z_{n},\bar{z}_{n})\frac{\bar{z}_{n-1,\tilde{j}}}{\bar{z}_{\tilde{j}n}\bar{z}_{n-1,n}}\\
+&\sum_{j>i=\tilde{j}}{z_{ij}^{2}}{\cal I}_{221}(z_{\tilde{j}},z_{j},\bar{z}_{j})\frac{\bar{z}_{1n}}{\bar{z}_{n\tilde{j}}\bar{z}_{1\tilde{j}}}
+\sum_{j>i}{z_{ij}^{2}}{\cal I}_{221}(z_{1},z_{j},\bar{z}_{1})\frac{\bar{z}_{\tilde{j},n-1}}{\bar{z}_{n-1,1}\bar{z}_{\tilde{j}1}}\\
+&{z_{n-1,n}^{2}}{\cal I}_{221}(z_{n-1},z_{n},\bar{z}_{n-1})\frac{\bar{z}_{n1}}{\bar{z}_{1,n-1}\bar{z}_{n,n-1}}\Big).
\ea
\ee
where ${\cal I}_{221}$ is defined by eq.(\ref{I1111}). In particular, one can take $n=4$\footnote{One should note that the coefficient of $\frac{\bar{\eta}_{\tilde{j}}}{f}\frac{\partial f}{\partial\bar{\eta}_{\tilde{j}}}$ in \cite{He:2019vzf} can be expressed as the linear combination of ${\cal I}_{221}$, and one can perform proper arrangements of  ${\cal I}_{221}$ to find that the above equation (\ref{secondTT}) is consistent with the coefficient $\mathcal{I}_{221111}$ given in \cite{He:2019vzf}. A similar situation happens in $G^{T\bar T}_3$.} to compare with the first order deformation of the four-point function given in \cite{He:2019vzf}. The third term of eq.(\ref{TTdeformation}) is the complex conjugate of the second term $G^{T\bar T}_2$. We will not repeat the details here.
The fourth term of eq.(\ref{TTdeformation}) is
\be
\ba\label{fourTT}
G^{T\bar T}_4=&\int d^{2}z\sum_{j,\tilde{j}=2}^{n-2}\frac{\eta_{j}\bar{\eta}_{\tilde{j}}}{f}\frac{\partial^{2}f}{\partial\bar{\eta}_{\tilde{j}}\partial\eta_{j}}\bar{\tilde{q}}\tilde{q}\\
=&\sum_{j,\tilde{j}=2}^{n-2}\frac{\eta_{j}\bar{\eta}_{\tilde{j}}}{f}\frac{\partial^{2}f}{\partial\bar{\eta}_{\tilde{j}}\partial\eta_{j}}\times\Big({\cal I}_{11}(z_{j},\bar{z}_{1})\frac{\bar{z}_{\tilde{j},n-1}}{\bar{z}_{n-1,1}\bar{z}_{\tilde{j}1}}\frac{z_{1n}}{z_{nj}z_{1j}}
+{\cal I}_{11}(z_{n-1},\bar{z}_{1})\frac{\bar{z}_{\tilde{j},n-1}}{\bar{z}_{n-1,1}\bar{z}_{\tilde{j}1}}\frac{z_{n1}}{z_{1,n-1}z_{n,n-1}}\\&+{\cal I}_{11}(z_{n},\bar{z}_{1})\frac{\bar{z}_{\tilde{j},n-1}}{\bar{z}_{n-1,1}\bar{z}_{\tilde{j}1}}\frac{z_{n-1,j}}{z_{jn}z_{n-1,n}}
+{\cal I}_{11}(z_{1},\bar{z}_{\tilde{j}})\frac{\bar{z}_{1n}}{\bar{z}_{n\tilde{j}}\bar{z}_{1\tilde{j}}}\frac{z_{j,n-1}}{z_{n-1,1}z_{j1}}\\&+{\cal I}_{11}(z_{j},\bar{z}_{\tilde{j}})\frac{\bar{z}_{1n}}{\bar{z}_{n\tilde{j}}\bar{z}_{1\tilde{j}}}\frac{z_{1n}}{z_{nj}z_{1j}}
+{\cal I}_{11}(z_{n-1},\bar{z}_{\tilde{j}})\frac{\bar{z}_{1n}}{\bar{z}_{n\tilde{j}}\bar{z}_{1\tilde{j}}}\frac{z_{n1}}{z_{1,n-1}z_{n,n-1}}\\&+{\cal I}_{11}(z_{n},\bar{z}_{\tilde{j}})\frac{\bar{z}_{1n}}{\bar{z}_{n\tilde{j}}\bar{z}_{1\tilde{j}}}\frac{z_{n-1,j}}{z_{jn}z_{n-1,n}}
+{\cal I}_{11}(z_{1},\bar{z}_{n-1})\frac{\bar{z}_{n1}}{\bar{z}_{1,n-1}\bar{z}_{n,n-1}}\frac{z_{j,n-1}}{z_{n-1,1}z_{j1}}\\&+{\cal I}_{11}(z_{j},\bar{z}_{n-1})\frac{\bar{z}_{n1}}{\bar{z}_{1,n-1}\bar{z}_{n,n-1}}\frac{z_{1n}}{z_{nj}z_{1j}}+{\cal I}_{11}(z_{n},\bar{z}_{n-1})\frac{\bar{z}_{n1}}{\bar{z}_{1,n-1}\bar{z}_{n,n-1}}\frac{z_{n-1,j}}{z_{jn}z_{n-1,n}}\\&
+{\cal I}_{11}(z_{1},\bar{z}_{n})\frac{\bar{z}_{n-1,\tilde{j}}}{\bar{z}_{\tilde{j}n}\bar{z}_{n-1,n}}\frac{z_{j,n-1}}{z_{n-1,1}z_{j1}}+{\cal I}_{11}(z_{j},\bar{z}_{n})\frac{\bar{z}_{n-1,\tilde{j}}}{\bar{z}_{\tilde{j}n}\bar{z}_{n-1,n}}\frac{z_{1n}}{z_{nj}z_{1j}}\\&
+{\cal I}_{11}(z_{n-1},\bar{z}_{n})\frac{\bar{z}_{n-1,\tilde{j}}}{\bar{z}_{\tilde{j}n}\bar{z}_{n-1,n}}\frac{z_{n1}}{z_{1,n-1}z_{n,n-1}}\Big).
\ea
\ee
where ${\cal I}_{11}$ is given by (\ref{I12}) after the regularization, and the terms associated with ${\cal I}_{11}(z_{1},\bar{z}_{1})$, ${\cal I}_{11}(z_{j},\bar{z}_{j})$, ${\cal I}_{11}(z_{n-1},\bar{z}_{n-1})$ and ${\cal I}_{11}(z_{n},\bar{z}_{n})$ in eq.(\ref{fourTT}) have been removed by renormalization due to logarithmic divergence. To finish the section, one can sum over these three terms,
\be\ba
\l\int_{C} d^2z\vev{T\bar{T}(z,\bar z)\phi_1(z_1)...\phi_n(z_n)}=\Big(G^{T\bar T}_1+2 \Re\Big(G^{T\bar T}_2\Big)+G^{T\bar T}_4\Big)\langle O_{n}\rangle
\ea\ee
 to obtain a first order correction to the $T\bar{T}$ deformed higher point correlation function. {As a consistency check\footnote{Note that the coefficient of $\frac{\eta_{j}\bar{\eta}_{\tilde{j}}}{f}\frac{\partial^{2}f}{\partial\bar{\eta}_{\tilde{j}}\partial\eta_{j}}$ in \cite{He:2019vzf} can be expressed by a linear combination of ${\cal I}_{11}$, and one can do proper arrangements of  ${\cal I}_{11}$ to find the above equation (\ref{fourTT}) coincides with coefficient $\mathcal{I}_{11111111}$ given in \cite{He:2019vzf}.}, one can follow the similar process\footnote{To compare with the first order deformation of the four-point function in  \cite{He:2019vzf}, our notation must be transferred into theirs in terms of eq.(\ref{notation}).} of taking $n=4$ to reproduce the first order deformation of the four-point correlation function given by eq.(12) in \cite{He:2019vzf}.}
\subsection{n-point correlation function in $J\bar{T}$ deformed CFTs}
In this subsection, we compute the n-point correlation function in the $J\bar{T}$ deformed CFTs. Since
\be
\ba
\langle TO_{n}\rangle=&\sum_{i=1}^{n}\frac{h}{(z-z_{i})^{2}}\langle O_{n}\rangle+\sum_{i=1}^{n}\sum_{j\neq i}\frac{a}{z-z_{i}}\frac{1}{z_{ji}}\langle O_{n}\rangle+\sum_{j=2}^{n-2}\eta_{j}\frac{\partial f(\eta_{j},\bar{\eta}_{j})}{\partial\eta_{j}}\times\\&\Big(\frac{1}{z-z_{1}}\frac{z_{j,n-1}}{z_{n-1,1}z_{j1}}+\frac{1}{z-z_{j}}\frac{z_{1n}}{z_{nj}z_{1j}}+\frac{1}{z-z_{n-1}}\frac{z_{n1}}{z_{1,n-1}z_{n,n-1}}+\frac{1}{z-z_{n}}\frac{z_{n-1,j}}{z_{jn}z_{n-1,n}}\Big){O_{L}O_{R}},
\ea
\ee
then the first order correction to the {$J\bar{T}$ deformed} correlation function is
\begin{align}
\langle J\bar{T}O_{n}\rangle=&\Big(\sum_{i=1}^{n}\frac{q_{i}}{z-z_{i}}\Big)
\sum_{j=1}^{n}\frac{\bar{h}}{(\bar{z}-\bar{z}_{j})^{2}}\langle O_{n}\rangle
+\Big(\sum_{i=1}^{n}\frac{q_{i}}{z-z_{i}}\Big)
\sum_{j=1}^{n}\sum_{k\neq j}\frac{\bar{a}}{\bar{z}-\bar{z}_{j}}\frac{1}{\bar{z}_{kj}}\langle O_{n}\rangle\nonumber\\
+&\Big(\sum_{i=1}^{n}\frac{q_{i}}{z-z_{i}}\Big)
\sum_{j=2}^{n-2}\bar{\eta_{j}}\frac{\partial f(\eta_{j},\bar{\eta}_{j})}{\partial\bar{\eta_{j}}}{O_{L}O_{R}}\nonumber\\
&\Big(\frac{1}{\bar{z}-\bar{z}_{1}}\frac{\bar{z}_{j,n-1}}{\bar{z}_{n-1,1}\bar{z}_{j1}}
+\frac{1}{\bar{z}-\bar{z}_{j}}\frac{\bar{z}_{1n}}{\bar{z}_{nj}\bar{z}_{1j}}+\frac{1}{\bar{z}-\bar{z}_{n-1}}
\frac{\bar{z}_{n1}}{\bar{z}_{1,n-1}\bar{z}_{n,n-1}}+\frac{1}{\bar{z}-\bar{z}_{n}}
\frac{\bar{z}_{n-1,j}}{\bar{z}_{jn}\bar{z}_{n-1,n}}\Big).
\end{align}
In terms of $\langle J\bar{T}O_{n}\rangle$, we have to integrate the above equation over the complex plane using a proper regularization procedure
\be
\ba\label{JT-npoint}
&\int d^{2}z \langle J\bar{T}O_{n}\rangle\\=& \sum_{i=1}^{n}\sum_{i\neq j}^{n} \frac{2\pi q_{i}\bar{h}}{\bar{z}_{ij}}\langle O_{n}\rangle+\sum_{i=1}^{n}\sum_{i\neq j}^{n}\sum_{j\neq k}^{n}\mathcal{I}_{i, j}\frac{q_{i}\bar{a}}{\bar{z}_{jk}}\langle O_{n}\rangle+ \sum_{j=2}^{n-2}\bar{\eta_{j}}\frac{\partial f(\eta_{j},\bar{\eta}_{j})}{\partial\bar{\eta_{j}}}\Big(\sum_{i=2}^{n}\mathcal{I}_{i,1}{q_{i}}\frac{\bar{z}_{j,n-1}}{\bar{z}_{n-1,1}\bar{z}_{j1}}\\
+&\sum_{i=1,i\neq j}^{n}\mathcal{I}_{i,j}{q_{i}}\frac{\bar{z}_{1n}}{\bar{z}_{nj}\bar{z}_{1j}}+\sum_{i=1,i\neq n-1}^{n}\mathcal{I}_{i,n-1}{q_{i}}\frac{\bar{z}_{n1}}{\bar{z}_{1,n-1}\bar{z}_{n,n-1}}+\sum_{i=1}^{n-1}\mathcal{I}_{i,n}{q_{i}}\frac{\bar{z}_{n-1,j}}{\bar{z}_{jn}\bar{z}_{n-1,n}}\Big){O_{L}O_{R}},
\ea
\ee
where $\mathcal{I}_{ij}$ is defined in Appendix \ref{sec:integral}. As a consistency check, one can take $n=4$ and apply the relations given by eq.(\ref{JTconvention}) to reproduce the first order $J\bar{ T}$ deformation of the four-point correlation function in CFTs given by \cite{He:2019vzf}.

\section{OTOC in the deformed Ising model}
The OTOC has been regarded as a diagnostic
of quantum chaos \cite{Roberts:2014ifa,Shenker:2014cwa,Maldacena:2015waa}. A field theory with gravity dual is
proposed to exhibit the maximal Lyapunov exponent that measures the growth rate
of the OTOC. In this section, we investigate the OTOC between pairs of operators $W, V$
$$
\frac{\langle W(t) V W(t) V\rangle_{\beta}}{\langle W(t) W(t)\rangle_{\beta}\langle V V\rangle_{\beta}}
$$
in the deformed CFTs to check whether the chaotic property is preserved or not after the
$T \bar{T}$ or $J\bar{T}$ deformation perturbatively. $\langle...\rangle_{\beta}$ is denoted by the correlation function on the cylinder. Since the OTOC can be broadly regarded as one of
the quantities characterizing the chaotic or integrable behavior, our current study will shed light
on the integrability/chaos after the $T \bar{T}$ or $J\bar{T}$ deformation.

In the thermal four-point correlators, $\langle\mathcal{O}(x, t) \cdots\rangle_{\beta}$, $, x, t$ are the coordinates of the spatially infinite thermal system\footnote{In particular, we use the 2D deformed theory on the cylinder.}, one can compute the vacuum expectation values through the conformal transformation
$$
\left\langle\mathcal{O}\left(x_{1}, t_{1}\right) \cdots\right\rangle_{\beta}=\left(\frac{2 \pi z_{1}}{\beta}\right)^{h}\left(\frac{2 \pi \bar{z}_{1}}{\beta}\right)^{h}\left\langle\mathcal{O}\left(z_{1}, \bar{z}_{1}\right) \cdots\right\rangle,
$$
where $z_{i}, \bar{z}_{i}$ are
\be\label{conformalmap}
z_{i}\left(x_{i}, t_{i}\right)=e^{\frac{2 \pi}{\beta}\left(x_{i}+t_{i}\right)}, \quad \bar{z}_{i}\left(x_{i}, t_{i}\right)=e^{\frac{2 \pi}{\beta}\left(x_{i}-t_{i}\right)}
\ee and $\langle...\rangle$ denotes the correlation function on the plane.
\subsection{OTOC in $T\bar{T}$-deformed Ising model}
In this subsection, we can use perturbations to calculate the $T\bar{T}$ deformation of OTOC in the Ising model \cite{DiFrancesco:1987ez}.
The first order $T\bar{T}$ deformation to the thermal correlator is the following
$$
\lambda \int d^{2} w\left\langle T \bar{T}(w, \bar{w}) \mathcal{O}\left(w_{1}, \bar{w}_{1}\right) \cdots\right\rangle_{\beta}
$$
where $w=x+t$ and $\bar{w}=x-t$ are coordinates on the cylinder.
To apply the $T \bar{T}$ deformed correlation function to the OTOC, we follow the steps
in \cite{He:2019vzf}, the first order deformed OTOC is
\be
\begin{split}\label{DeformedTT}
&C_{WV}(t)=\frac{\langle W(w_1,\bar{w}_1)W(w_2,\bar{w}_2)V(w_3,\bar{w}_3)V(w_4,\bar{w}_4)\rangle}{\langle W(w_1,\bar{w}_1)W(w_2,\bar{w}_2)\rangle\langle V(w_3,\bar{w}_3)V(w_4,\bar{w}_4)\rangle}\\
&\times\Big{(}1-\lambda(\frac{2\pi}{\beta})^2\int d^2z|z|^2\frac{\langle(T(z)-\frac{c}{24z^2})(\bar{T}(\bar{z})-\frac{c}{24\bar{z}^2})W(z_1,\bar{z}_1)W(z_2,\bar{z}_2)\rangle}{\langle W(z_1,\bar{z}_1)W(z_2,\bar{z}_2)\rangle}\\
&-\lambda(\frac{2\pi}{\beta})^2\int d^2z|z|^2\frac{\langle(T(z)-\frac{c}{24z^2})(\bar{T}(\bar{z})-\frac{c}{24\bar{z}^2})V(z_3,\bar{z}_3)V(z_4,\bar{z}_4)\rangle}{\langle V(z_3,\bar{z}_3)V(z_4,\bar{z}_4)\rangle}\\
&+\lambda(\frac{2\pi}{\beta})^2\int d^2z|z|^2\frac{\langle(T(z)-\frac{c}{24z^2})(\bar{T}(\bar{z})-\frac{c}{24\bar{z}^2})W(z_1,\bar{z}_1)W(z_2,\bar{z}_2)V(z_3,\bar{z}_3)V(z_4,\bar{z}_4)\rangle}{\langle W(z_1,\bar{z}_1)W(z_2,\bar{z}_2)V(z_3,\bar{z}_3)V(z_4,\bar{z}_4)\rangle}\\&+\mathcal{O}(\lambda^2)\Big{)}.
\end{split}
\ee

For generic two-dimensional CFTs, the four-point function on the plane is
\be\label{WVWV}
\langle W(z_1,\bar{z}_1)W(z_2,\bar{z}_2)V(z_3,\bar{z}_3)V(z_4,\bar{z}_4)\rangle=\frac{1}{z_{12}^{2h_w}z_{34}^{2h_v}}\frac{1}{\bar{z}_{12}^{2h_w}\bar{z}_{34}^{2h_v}}G(\eta,\bar{\eta})\,,
\ee
where $G(\eta,\bar{\eta})$ is associated with the conformal block.
In the Ising model, there are three types of $G(\eta,\bar{\eta})$, which are associated with the three Virasoro primary operators, e.g. identity operator $I$, spin operator $\sigma$, and  energy operator $\epsilon$. They are
\be\label{sigmasigma}
G_{\sigma\sigma}(\eta,\bar{\eta})=\frac{1}{2}\Big{|}\frac{1}{1-\eta}\Big{|}^{1/4}\big{(}|1+\sqrt{1-\eta}|+|1-\sqrt{1-\eta}|\big{)}\,,
\ee
\be\label{sigmaepsilon}
G_{\sigma\epsilon}(\eta,\bar{\eta})=\Big{|}\frac{2-\eta}{2\sqrt{1-\eta}}\Big{|}^2\,,
\ee
\be\label{epsilonepsilon}
G_{\epsilon\epsilon}(\eta,\bar{\eta})=\Big{|}\frac{1-\eta+\eta^2}{1-\eta}\Big{|}^2\,,
\ee
corresponding to $\langle\sigma\sigma\sigma\sigma\rangle$, $\langle\sigma\epsilon\sigma\epsilon\rangle$, and $\langle\epsilon\epsilon\epsilon\epsilon\rangle$ respectively \cite{DiFrancesco:1987ez}. Here $|f(\eta)|=\sqrt{f(\eta)}\sqrt{f(\bar{\eta})}$.

Then, the first order deformation can be calculated by considering different forms of $G(\eta,\bar{\eta})$. Here we take $\langle\sigma\sigma\sigma\sigma\rangle$ as an example, and the first order $T\bar{T}$ deformed OTOC (\ref{DeformedTT}) can be divided into three terms in terms of different powers of the central charge\footnote{The central charge $c$ is ${1\over 2}$ in the 2D Ising model.} $c$. The term with power $c^2$ is independent of $\delta C_{\sigma\sigma}$ such that
\be
\begin{split}
\delta C_{\sigma\sigma}\big{(}t, c^2\big{)}=&\lambda \frac{c^2}{24^2}(\frac{2\pi}{\beta})^2\int d^2z |z|^{-2}\\
=&\lambda \frac{c^2}{24^2}(\frac{2\pi}{\beta})^2 2\pi\int_{\frac{1}{\tilde{\Lambda}}}^\Lambda d^2\rho \frac{1}{\rho}\\
=&-\lambda \frac{c^2}{24^2}(\frac{2\pi}{\beta})^2 2\pi {\rm log}(\Lambda\tilde{\Lambda})\,.
\end{split}
\ee
Since this term is only associated with the logarithmic divergence, it can be regulated by the regularization procedure and it does not contribute to the OTOC.
Putting eqs.(\ref{WVWV}) and (\ref{sigmasigma}) into eq.(\ref{DeformedTT}), the term with $c^1$ of the $T\bar{T}$ deformed OTOC (\ref{DeformedTT}) is as follows
\be
\begin{split}
&\delta C_{\sigma\sigma}\big{(}t,c^1\big{)}=\frac{c\lambda}{24}(\frac{2\pi}{\beta})^2\,2\pi\,\\
&\times\Bigg{[}\frac{\eta\partial_\eta G(\eta,\bar{\eta})}{G(\eta,\bar{\eta})}z_{13}z_{24}\Big{(}\frac{z_1}{z_{12}z_{13}z_{14}}{\rm log}\frac{1}{|z_1|}-\frac{z_2}{z_{12}z_{23}z_{24}}{\rm log}\frac{1}{|z_2|}+\frac{z_3}{z_{13}z_{23}z_{34}}{\rm log}\frac{1}{|z_3|}-\frac{z_4}{z_{14}z_{24}z_{34}}{\rm log}\frac{1}{|z_4|}\Big{)}\\
&-\frac{\bar{\eta}\partial_{\bar{\eta}} G(\eta,\bar{\eta})}{G(\eta,\bar{\eta})}\bar{z}_{13}\bar{z}_{24}\Big{(}\frac{\bar{z}_1}{\bar{z}_{12}\bar{z}_{13}\bar{z}_{14}}{\rm log}|z_1|-\frac{\bar{z}_2}{\bar{z}_{12}\bar{z}_{23}\bar{z}_{24}}{\rm log}|z_2|+\frac{\bar{z}_3}{\bar{z}_{13}\bar{z}_{23}\bar{z}_{34}}{\rm log}|z_3|-\frac{\bar{z}_4}{\bar{z}_{14}\bar{z}_{24}\bar{z}_{34}}{\rm log}|z_4|\Big{)}\Bigg{]}\,.\label{dC1}
\end{split}
\ee

To apply the $T \bar{T}$-deformed correlation function to the OTOC, we follow the steps
in \cite{Roberts:2014ifa,He:2019vzf} to evaluate the OTOC by using the analytic continuation of the Euclideans of the four-point
function, writing
\be\label{OTOCcordinates}
\begin{split}
&z_1=e^{\frac{2\pi}{\beta}i\epsilon_1}\,, \bar{z}_1=e^{-\frac{2\pi}{\beta}i\epsilon_1}\,,\\
&z_2=e^{\frac{2\pi}{\beta}i\epsilon_2}\,, \bar{z}_2=e^{-\frac{2\pi}{\beta}i\epsilon_2}\,,\\
&z_3=e^{\frac{2\pi}{\beta}(t+i\epsilon_3-x)}\,, \bar{z}_3=e^{-\frac{2\pi}{\beta}(-t-i\epsilon_3-x)}\,,\\
&z_4=e^{\frac{2\pi}{\beta}(t+i\epsilon_4-x)}\,, \bar{z}_4=e^{-\frac{2\pi}{\beta}(-t-i\epsilon_4-x)}\,,
\end{split}
\ee
and substituting $\epsilon_1=0$, $\epsilon_2=\epsilon_1+\beta/2$, $\epsilon_4=\epsilon_3+\beta/2$ into \eqref{dC1}. To get the late time behavior of the OTOC, one can expand around $e^{-\frac{2\pi t}{\beta}}$
\be\label{TTOTOC1}
\delta C_{\sigma\sigma}\big{(}t, c^1\big{)}=\frac{c\lambda\pi^4 x e^{-\frac{4\pi(i\epsilon_3+x)}{\beta}}(e^{\frac{8\pi x}{\beta}}-1)}{3\beta^3}(e^{-\frac{2\pi t}{\beta}})^2+\mathcal{O}\Big{(}(e^{-\frac{2\pi t}{\beta}})^3\Big{)}\,,
\ee

Finally, we can apply a similar approach to calculate the order $c^0$ part,
\be\label{TTOTOC2}
\begin{split}
\delta C_{\sigma\sigma}\big{(}t, c^0\big{)}=&\lambda(\frac{2\pi}{\beta})^2\int d^2z|z|^2\Big{[}\frac{\langle T(z)\bar{T}(\bar{z})W(z_1,\bar{z}_1)W(z_2,\bar{z}_2)V(z_3,\bar{z}_3)V(z_4,\bar{z}_4)\rangle}{\langle W(z_1,\bar{z}_1)W(z_2,\bar{z}_2)V(z_3,\bar{z}_3)V(z_4,\bar{z}_4)\rangle}\\
-&\frac{\langle T(z)\bar{T}(\bar{z})W(z_1,\bar{z}_1)W(z_2,\bar{z}_2)\rangle}{\langle W(z_1,\bar{z}_1)W(z_2,\bar{z}_2)\rangle}-\frac{\langle T(z)\bar{T}(\bar{z})V(z_3,\bar{z}_3)V(z_4,\bar{z}_4)\rangle}{\langle V(z_3,\bar{z}_3)V(z_4,\bar{z}_4)\rangle}\Big{]}\\
=&\frac{4\pi^3\lambda h_w e^{-\frac{4\pi(i e_3+x)}{\beta}}(-1-4e^{\frac{2\pi x}{\beta}}+4e^{\frac{6\pi x}{\beta}}+e^{\frac{8\pi x}{\beta}})}{\beta^2\epsilon_0^2}(e^{-\frac{2\pi t}{\beta}})^2+\mathcal{O}\Big{(}(e^{-\frac{2\pi t}{\beta}})^3\Big{)}\,,
\end{split}
\ee
where $\epsilon_0$ is a cutoff denoted by $|z_i|^2=z_i\bar{z}_i+\epsilon_0^2$.

Following similar steps, one can obtain the OTOC associated with  $\langle\sigma\epsilon\sigma\epsilon\rangle_\beta$ and $\langle\epsilon\epsilon\epsilon\epsilon\rangle_\beta$, respectively. The first order corrections to the OTOC $\langle\sigma\epsilon\sigma\epsilon\rangle_\beta$ contain the following three individual contributions in the late time limit
\be\label{TTOTOC3}
\delta C_{\sigma\epsilon}\big{(}t, c^2\big{)}=-\frac{c^2\lambda\pi^3{\rm log}(\Lambda\tilde{\Lambda})}{72\beta^2}\,,
\ee
\be\label{TTOTOC4}
\delta C_{\sigma\epsilon}\big{(}t, c^1\big{)}= \frac{8c\lambda\pi^4 x e^{-\frac{4\pi(i\epsilon_3+x)}{\beta}}(e^{\frac{8\pi x}{\beta}}-1)}{3\beta^3}(e^{-\frac{2\pi t}{\beta}})^2+\mathcal{O}\Big{(}(e^{-\frac{2\pi t}{\beta}})^3\Big{)}\,,
\ee
\be\label{TTOTOC5}
\delta C_{\sigma\epsilon}\big{(}t, c^0\big{)}= \frac{32\lambda\pi^3 h_w e^{-\frac{4\pi(i e_3+x)}{\beta}}(e^{\frac{8\pi x}{\beta}}-1)}{\beta^2\epsilon_0^2}(e^{-\frac{2\pi t}{\beta}})^2+\mathcal{O}\Big{(}(e^{-\frac{2\pi t}{\beta}})^3\Big{)}\,.
\ee

The first order corrections to the OTOC $\langle\epsilon\epsilon\epsilon\epsilon\rangle_\beta$ contain the following three terms in the late time limit
\be\label{TTOTOC6}
\delta C_{\epsilon\epsilon}\big{(}t, c^2\big{)}= -\frac{c^2\lambda\pi^3{\rm log}(\Lambda\tilde{\Lambda})}{72\beta^2}\,,
\ee
\be\label{TTOTOC7}
\delta C_{\epsilon\epsilon}\big{(}t, c^1\big{)}= \frac{64c\lambda\pi^4 x e^{-\frac{4\pi(i\epsilon_3+x)}{\beta}}(e^{\frac{8\pi x}{\beta}}-1)}{3\beta^3}(e^{-\frac{2\pi t}{\beta}})^2+\mathcal{O}\Big{(}(e^{-\frac{2\pi t}{\beta}})^3\Big{)}\,,
\ee
\be\label{TTOTOC8}
\delta C_{\epsilon\epsilon}\big{(}t, c^0\big{)}= \frac{256\lambda\pi^3 h_w e^{-\frac{4\pi(i e_3+x)}{\beta}}(e^{\frac{8\pi x}{\beta}}-1)}{\beta^2\epsilon_0^2}(e^{-\frac{2\pi t}{\beta}})^2+\mathcal{O}\Big{(}(e^{-\frac{2\pi t}{\beta}})^3\Big{)}\,.
\ee
In these examples, one can see the late time limit of OTOC (\ref{TTOTOC1})-(\ref{TTOTOC8}) associated with one pair of the operators is not changed, up to the first order $T\bar{T}$ deformation. In this sense, the $T\bar{T}$ deformation preserves the integrable property of the un-deformed Ising model.
\subsection{OTOC in $J\bar{T}$-deformed Ising model}
Similarly, one can calculate the OTOC in the $J\bar{T}$-deformed Ising model. The first order $J\bar{T}$ deformation of the thermal correlator is as follows
$$
\lambda \int d^{2} w\left\langle J \bar{T}(w, \bar{w}) \mathcal{O}\left(w_{1}, \bar{w}_{1}\right) \cdots\right\rangle_{\beta}.
$$
where $w=x+t$ and $\bar{w}=x-t$ are coordinates on the cylinder, which is similar to the setup in the above subsection.
For the $J\bar{T}$ deformation, one has to replace the $T$ operator in eq.(\ref{DeformedTT}) with the conserved current $J$ to obtain the deformed OTOC
\be
\begin{split}\label{OTOCJT}
\tilde{C}_{WV}(t)=&\frac{\langle W(w_1,\bar{w}_1)W(w_2,\bar{w}_2)V(w_3,\bar{w}_3)V(w_4,\bar{w}_4)\rangle}{\langle W(w_1,\bar{w}_1)W(w_2,\bar{w}_2)\rangle\langle V(w_3,\bar{w}_3)V(w_4,\bar{w}_4)\rangle}\\
&\times\Big{[}1-\lambda\int d^2z\frac{2\pi\bar{z}}{\beta}\frac{\langle J(\bar{T}(\bar{z})-\frac{c}{24\bar{z}^2})W(z_1,\bar{z}_1)W(z_2,\bar{z}_2)\rangle}{\langle W(z_1,\bar{z}_1)W(z_2,\bar{z}_2)\rangle}\\
&-\lambda\int d^2z\frac{2\pi\bar{z}}{\beta}\frac{\langle J(\bar{T}(\bar{z})-\frac{c}{24\bar{z}^2})V(z_3,\bar{z}_3)V(z_4,\bar{z}_4)\rangle}{\langle V(z_3,\bar{z}_3)V(z_4,\bar{z}_4)\rangle}\\
&+\lambda\int d^2z\frac{2\pi\bar{z}}{\beta}\frac{\langle J(\bar{T}(\bar{z})-\frac{c}{24\bar{z}^2})W(z_1,\bar{z}_1)W(z_2,\bar{z}_2)V(z_3,\bar{z}_3)V(z_4,\bar{z}_4)\rangle}{\langle W(z_1,\bar{z}_1)W(z_2,\bar{z}_2)V(z_3,\bar{z}_3)V(z_4,\bar{z}_4)\rangle}+\mathcal{O}(\lambda^2)\Big{]}\,,
\end{split}
\ee
where $w_1, w_2,w_3,w_4$ are the operator positions on the cylinder and $z_1, z_2,z_3,z_4$ are the corresponding coordinates on the plane, and the map between the $w-$ plane and $z-$plane is given by eq.(\ref{conformalmap}).
The first order correction to the OTOC is following
\be\label{JTbarWV}
\begin{split}
\delta\tilde{C}_{WV}(t)&=\lambda\int d^2z\frac{2\pi\bar{z}}{\beta}\Big{[}(\frac{q_3}{z-z_3}+\frac{q_4}{z-z_4})h_w\frac{\bar{z}_{12}^2}{(\bar{z}-\bar{z}_1)^2(\bar{z}-\bar{z}_2)^2}\\&+
(\frac{q_1}{z-z_1}+\frac{q_2}{z-z_2})h_v\frac{\bar{z}_{34}^2}{(\bar{z}-\bar{z}_3)^2(\bar{z}-\bar{z}_4)^2}\\
&+\Big{(}\sum_{i=1}^4\frac{q_i}{z-z_i}\Big{)}\frac{\bar{z}_{14}\bar{z}_{23}}{\prod_{i=1}^4(\bar{z}-\bar{z}_i)}\frac{\bar{\eta}\partial_{\bar{\eta}}G(\eta,\bar{\eta})}{G(\eta,\bar{\eta})}\Big{]}\,.
\end{split}
\ee

One can apply a similar approach to that in the above subsection and impose the equations (\ref{OTOCcordinates}), (\ref{sigmasigma}), (\ref{sigmaepsilon}), and (\ref{epsilonepsilon}) to eq.(\ref{JTbarWV}). Finally, one can expand around $e^{-\frac{2\pi t}{\beta}}$ to obtain the late time behavior of the first order $J\bar{T}$ deformation to the OTOC as follows
\be\label{sigmaepsilonJT1}
\begin{split}
\delta \tilde{C}_{\sigma\sigma}(t)=&-\frac{\lambda\pi^2(q_1+q_2)e^{-\frac{2\pi i \epsilon_3}{\beta}}}{2\beta\epsilon_0}(e^{-\frac{2\pi t}{\beta}})\\-&\frac{\lambda\pi^2(q_1+q_2)e^{-\frac{4\pi (i \epsilon_3+x)}{\beta}}(e^{\frac{4\pi x}{\beta}}-1)}{2\beta\epsilon_0}(e^{-\frac{2\pi t}{\beta}})^2+\mathcal{O}\Big{(}(e^{-\frac{2\pi t}{\beta}})^3\Big{)}\,,
\end{split}
\ee
\be\label{sigmaepsilonJT2}
\delta \tilde{C}_{\sigma\epsilon}(t)=\delta \tilde{C}_{\epsilon\epsilon}(t)=-\frac{32\lambda\pi^2(q_1+q_2)e^{-\frac{4\pi(i\epsilon_3+x)}{\beta}}}{\beta\epsilon_0}(e^{-\frac{2\pi t}{\beta}})^2+\mathcal{O}\Big{(}(e^{-\frac{2\pi t}{\beta}})^3\Big{)}\,.
\ee
Up to the first order $J\bar{T}$ deformation, one can see that the late time limit of OTOC (\ref{sigmaepsilonJT1})(\ref{sigmaepsilonJT2}) associated with different operators is not changed.

{To end this section, we extract the exact large-time behavior of the $T\bar{T}$ and $J\bar{T}$ deformed Ising model to confirm that these deformations preserve the integrable property of the un-deformed Ising model. Since the spectrum in the Ising model is finite-dimensional and the c central charge is one-half, one can analytically extract the first order deformation of the OTOC in the deformed theories without the large c expansion. From a quantum information point of view, our investigations support that the TTbar/JTbar deformation preserves the integrable properties of the original theory up to the first order deformation. In a phenomenological sense, it is a good check that the OTOC is a good quantity for investigating quantum chaos and quantum integrability \cite{Smirnov:2016lqw}. }
\section{Conclusions and discussions}
In this paper, we apply a perturbative CFT approach to calculate the first order correction to the generic n-point correlation function in $T\bar{T}$ and $J\bar{T}$ deformed CFTs, following the approach in \cite{He:2019vzf,He:2019ahx,He:2020udl}. Since conformal symmetry can be regarded as an approximate symmetry to the CFTs in the first order $T\bar{T}$ and $J\bar{T}$ deformations, one can make use of the conformal Ward identities to construct the first order deformation of the n-point correlation function in CFTs. Since the OTOC has been regarded as a diagnostic of quantum chaos \cite{Roberts:2014ifa,Shenker:2014cwa,Maldacena:2015waa} and its late time behavior characterizes the quantum chaos signal of the theory, we calculate the late time behavior of the OTOC in the $T\bar{T}$ and $J\bar{T}$ deformed Ising model. It turns out that the late time limit of the OTOC is not changed by the deformations, and the physical situation is similar to that of the one dimensional $T\bar{T}$ deformation of the $\text{SYK}_4$ model \cite{Gross:2019uxi}. That is to say the $T\bar{T}$ and $J\bar{T}$ preserve the integrable property of the un-deformed 2D Ising model in terms of the late time limit of the OTOC. This is an apparent evidence that deforming a theory by quadratic composites of KdV currents \cite{LeFloch:2019wlf} preserves these un-deformed symmetries. As suggested in \cite{LeFloch:2019wlf}, the important property of these irrelevant deformations is that they preserve many symmetries of the un-deformed theory: any current whose charge commutes with the charges of the currents building the deformation can be adjusted so that the current remains conserved in the new theory. Our investigation supports the statement by probing the OTOC in $T\bar{T}$ and $J\bar{T}$ deformed theories.

The first-order corrections to the higher point correlation functions of the generic TTbar/JTbar deformed CFTs offered are very useful to probe quantum chaos and quantum entanglement. Since the conformal symmetry still holds approximately at the first order deformation, one can apply for the replica trick to investigate several quantum information quantities, e.g., higher-order Renyi entropy, entanglement purification, quantum teleportation, entanglement negativity, and higher-order out-of-time-ordered correlation (OTOC) functions in terms of replica tricks in 1+1 dimensional CFTs. All these quantum information quantities give us deep insights into quantum chaos and quantum entanglement. In generic QFTs, it is not possible to apply the replica trick, and so it is natural to study these quantities associated with quantum entanglement in QFT. Furthermore, to investigate how these quantities capture the essential properties of quantum chaos or quantum entanglement, one has to apply the higher-point correlation functions to learn more about these deformed theories. In this sense, we can offer field theory data in the TTbar/JTbar deformed theory, which are potentially helpful for understanding quantum chaos and entanglement of TTbar/JTbar deformed theory.

\section*{Acknowledgments}
We would like to thank Jiahui Bao, Bin Chen, Yi-hong Gao, Miao He, Chris Lau, Li Li, Yi Li, Yunfeng Jiang, Hao Ouyang, Hongfei Shu, Hao-Yu Sun, Yuan Sun, Stefan Theisen, and Yu-Xuan Zhang for their useful discussions related to this work. We would like to thank APCTP for their online hospitality in a workshop, ``$T\bar{T}$ deformation and Integrability’’ and the participants for their valuable comments. We would also like to acknowledge the financial support from Jilin University and the Max Planck Partner group, as well as from Natural Science Foundation of China Grant Numbers 12075101 and 12047569.

\appendix
\section{Simplified notations}
\subsection{Simplifying $G^{T\bar T}_1$}
The eq.(\ref{first-term}) can be expressed as following individual terms
\be
\ba
G^{T\bar T}_1&=\int d^{2}z(F+p)(\bar{F}+\bar{p})
\\=&\Big(\frac{h}{n-1}\Big)^2\sum_{i=1}^{n}\sum_{j>i}\sum_{\tilde{i}=1,\tilde{i}\neq i,j}^{n}\sum_{\tilde{j}>\tilde{i},\tilde{j}\neq i,j}\int d^{2}z\frac{z_{ij}^{2}}{(z-z_{i})^2(z-z_{j})^2}\frac{\bar{z}_{\tilde{i}\tilde{j}}^{2}}{(\bar{z}-\bar{z}_{\tilde{i}})^2(\bar{z}-\bar{z}_{\tilde{j}})^2}\Big|_{\tilde{i}\neq i,j,\tilde{j}\neq i,j,j>i,\tilde{j}>\tilde{i}}\\+&\Big(\frac{h}{n-1}\Big)^2\sum_{i=1}^{n}\sum_{j>i}\sum_{\tilde{i}=1,\tilde{i}\neq i,j}^{n}\int d^{2}z\frac{z_{ij}^{2}}{(z-z_{i})^2(z-z_{j})^2}\frac{\bar{z}_{\tilde{i}\tilde{j}}^{2}}{(\bar{z}-\bar{z}_{\tilde{i}})^2(\bar{z}-\bar{z}_{j})^2}\Big|_{\tilde{j}=j,\tilde{i}\neq i,j,j>i,\tilde{i}}\\+&\Big(\frac{h}{n-1}\Big)^2\sum_{i=1}^{n}\sum_{j>i}\sum_{\tilde{i}=1,\tilde{i}\neq i,j}^{n}\int d^{2}z\frac{z_{ij}^{2}}{(z-z_{i})^2(z-z_{j})^2}\frac{\bar{z}_{\tilde{i}\tilde{j}}^{2}}{(\bar{z}-\bar{z}_{\tilde{i}})^2(\bar{z}-\bar{z}_{i})^2}\Big|_{\tilde{i}\neq i,j,\tilde{j}=i,j>i,j>\tilde{i}}\\+&\Big(\frac{h}{n-1}\Big)^2\sum_{i=1}^{n}\sum_{j>i}\sum_{\tilde{j}>\tilde{i}}\int d^{2}z\frac{z_{ij}^{2}}{(z-z_{i})^2(z-z_{j})^2}\frac{\bar{z}_{\tilde{i}\tilde{j}}^{2}}{(\bar{z}-\bar{z}_{j})^2(\bar{z}-\bar{z}_{\tilde{j}})^2}\Big|_{\tilde{i}\neq i,\tilde{i}=j,j>i,\tilde{j}>j>i}\\+&\Big(\frac{h}{n-1}\Big)^2\sum_{i=1}^{n}\sum_{j>i}\sum_{\tilde{j}>i,\tilde{j}\neq j}\int d^{2}z\frac{z_{ij}^{2}}{(z-z_{i})^2(z-z_{j})^2}\frac{\bar{z}_{\tilde{i}\tilde{j}}^{2}}{(\bar{z}-\bar{z}_{i})^2(\bar{z}-\bar{z}_{\tilde{j}})^2}\Big|_{\tilde{i}=i,j,\tilde{j}\ne j,j>i,\tilde{j}>i}\\+&\Big(\frac{h}{n-1}\Big)^2\sum_{i=1}^{n}\sum_{j>i}\int d^{2}z\frac{z_{ij}^{2}}{(z-z_{i})^2(z-z_{j})^2}\frac{\bar{z}_{\tilde{i}\tilde{j}}^{2}}{(\bar{z}-\bar{z}_{i})^2(\bar{z}-\bar{z}_{j})^2}\Big|_{\tilde{i}=i,\tilde{j}=j,j>i,j>i}.
\ea
\ee
Using the integral notation in Appendix \ref{sec:integral}, $\int d^{2}z(F+p)(\bar{F}+\bar{p})$ can be rewritten as
\be
\ba\label{firstTT}
G^{T\bar T}_1&=\int d^{2}z(F+p)(\bar{F}+\bar{p})\\=&\Big(\frac{h}{n-1}\Big)^2\sum_{i=1}^{n}\sum_{j>i}\sum_{\tilde{i}=1,\tilde{i}\neq i,j}^{n}\sum_{\tilde{j}>\tilde{i},\tilde{j}\neq i,j}z_{ij}^{2}\bar{z}_{\tilde{i}\tilde{j}}^{2}{\cal I}_{2222}(z_{i},z_{j},\bar{z}_{\tilde{i}},\bar{z}_{\tilde{j}})\Big|_{\tilde{i}\neq i,j,\tilde{j}\neq i,j,j>i,\tilde{j}>\tilde{i}}\\+&\Big(\frac{h}{n-1}\Big)^2\sum_{i=1}^{n}\sum_{j>i}\sum_{\tilde{i}=1,\tilde{i}\neq i,j}^{n}z_{ij}^{2}\bar{z}_{\tilde{i}j}^{2}{\cal I}_{2222}(z_{i},z_{j},\bar{z}_{\tilde{i}},\bar{z}_{j})\Big|_{\tilde{j}=j,\tilde{i}\neq i,j,j>i,\tilde{i}}\\+&\Big(\frac{h}{n-1}\Big)^2\sum_{i=1}^{n}\sum_{j>i}\sum_{\tilde{i}=1,\tilde{i}\neq i,j}^{n}z_{ij}^{2}\bar{z}_{\tilde{i}i}^{2}{\cal I}_{2222}(z_{i},z_{j},\bar{z}_{i},\bar{z}_{\tilde{i}})\Big|_{\tilde{i}\neq i,j,\tilde{j}=i,j>i,i>\tilde{i}}\\+&\Big(\frac{h}{n-1}\Big)^2\sum_{i=1}^{n}\sum_{j>i}\sum_{\tilde{j}>\tilde{i}}z_{ij}^{2}\bar{z}_{j\tilde{j}}^{2}{\cal I}_{2222}(z_{j},z_{i},\bar{z}_{j},\bar{z}_{\tilde{j}})\Big|_{\tilde{i}\neq i,\tilde{i}=j,j>i,\tilde{j}>j>i}\\+&\Big(\frac{h}{n-1}\Big)^2\sum_{i=1}^{n}\sum_{j>i}\sum_{\tilde{j}>i,\tilde{j}\neq j}z_{ij}^{2}\bar{z}_{i\tilde{j}}^{2}{\cal I}_{2222}(z_{i},z_{j},\bar{z}_{i},\bar{z}_{\tilde{j}})\Big|_{\tilde{i}=i,j,\tilde{j}\ne j,j>i,\tilde{j}>i}\\+&\Big(\frac{h}{n-1}\Big)^2\sum_{i=1}^{n}\sum_{j>i}z_{ij}^{2}\bar{z}_{ij}^{2}{\cal I}_{2222}(z_{i},z_{j},\bar{z}_{i},\bar{z}_{j})\Big|_{\tilde{i}=i,\tilde{j}=j,j>i,j>i},
\ea
\ee
where ${\cal I}_{2222}$\footnote{This integral also appeared in the first order deformation of the four-point function given by \cite{He:2019vzf}. Here the applied regularization process is the same as the one used in \cite{He:2019vzf}.} is given by eq.(\ref{I1111}).

\subsection{Simplifying $G^{T\bar T}_2$}
The second term of eq.(\ref{TTdeformation}) is
\be
\ba
G^{T\bar T}_2&=\int d^{2}z(F+p)\sum_{\tilde{j}=2}^{n-2}\frac{\bar{\eta}_{\tilde{j}}}{f}\frac{\partial f}{\partial\bar{\eta}_{\tilde{j}}}\bar{\tilde{q}}\\
=&\frac{h}{n-1}\int d^{2}z\sum_{\tilde{j}=2}^{n-2}\frac{\bar{\eta}_{\tilde{j}}}{f}\frac{\partial f}{\partial\bar{\eta}_{\tilde{j}}}\Big(\sum_{i=1,i\neq1}^{n}\sum_{j>i}\frac{z_{ij}^{2}}{(z-z_{i})^2(z-z_{j})^2}\frac{1}{\bar{z}-\bar{z}_{1}}\frac{\bar{z}_{\tilde{j},n-1}}{\bar{z}_{n-1,1}\bar{z}_{\tilde{j}1}}\\
+&\sum_{i=1,i\neq\tilde{j}}^{n}\sum_{j>i,j\neq\tilde{j}}\frac{z_{ij}^{2}}{(z-z_{i})^2(z-z_{j})^2}\frac{1}{\bar{z}-\bar{z}_{\tilde{j}}}\frac{\bar{z}_{1,n}}{\bar{z}_{n,\tilde{j}}\bar{z}_{1\tilde{j}}}
+\sum_{i=1,i\neq\tilde{j}}^{n}\sum_{j=\tilde{j}>i}\frac{z_{ij}^{2}}{(z-z_{i})^2(z-z_{\tilde{j}})^2}\frac{1}{\bar{z}-\bar{z}_{\tilde{j}}}\frac{\bar{z}_{1n}}{\bar{z}_{n\tilde{j}}\bar{z}_{1\tilde{j}}}\\
+&\sum_{i=1,i\neq n-1}^{n}\sum_{j>i,j\neq n-1}\frac{z_{ij}^{2}}{(z-z_{i})^2(z-z_{j})^2}\frac{1}{\bar{z}-\bar{z}_{n-1}}\frac{\bar{z}_{n}-\bar{z}_{1}}{\bar{z}_{1,n-1}\bar{z}_{n,n-1}}\\+&\sum_{i=1,i\neq n-1}^{n}\sum_{j=n-1>i}\frac{z_{i,n-1}^{2}}{(z-z_{i})^2(z-z_{n-1})^2}\frac{1}{\bar{z}-\bar{z}_{n-1}}\frac{\bar{z}_{n}-\bar{z}_{1}}{\bar{z}_{1,n-1}\bar{z}_{n,n-1}}\\+&\sum_{i=1}^{n}\sum_{j>i,j\neq n}\frac{z_{ij}^{2}}{(z-z_{i})^2(z-z_{j})^2}\frac{1}{\bar{z}-\bar{z}_{n}}\frac{\bar{z}_{n-1,\tilde{j}}}{\bar{z}_{\tilde{j}n}\bar{z}_{n-1,n}}
+\sum_{i=1}^{n}\sum_{j=n>i}\frac{z_{in}^{2}}{(z-z_{i})^2(z-z_{n})^2}\frac{1}{\bar{z}-\bar{z}_{n}}\frac{\bar{z}_{n-1,\tilde{j}}}{\bar{z}_{\tilde{j}n}\bar{z}_{n-1,n}}\\
+&\sum_{j>i=\tilde{j}}\frac{z_{ij}^{2}}{(z-z_{\tilde{j}})^2(z-z_{j})^2}\frac{1}{\bar{z}-\bar{z}_{\tilde{j}}}\frac{\bar{z}_{1n}}{\bar{z}_{n\tilde{j}}\bar{z}_{1\tilde{j}}}
+\sum_{j>i}\frac{z_{ij}^{2}}{(z-z_{1})^2(z-z_{j})^2}\frac{1}{\bar{z}-\bar{z}_{1}}\frac{\bar{z}_{\tilde{j},n-1}}{\bar{z}_{n-1,1}\bar{z}_{\tilde{j}1}}\\
+&\frac{z_{n-1,n}^{2}}{(z-z_{n-1})^2(z-z_{n})^2}\frac{1}{\bar{z}-\bar{z}_{n-1}}\frac{\bar{z}_{n1}}{\bar{z}_{1,n-1}\bar{z}_{n,n-1}}\Big).
\ea
\ee
By using the integral notation, we obtain
\be
\ba\label{secondTT}
G^{T\bar T}_2 =&\int d^{2}z(F+p)\sum_{\tilde{j}=2}^{n-2}\frac{\bar{\eta}_{\tilde{j}}}{f}\frac{\partial f}{\partial\bar{\eta}_{\tilde{j}}}\bar{\tilde{q}}\\
=&\frac{h}{n-1}\sum_{\tilde{j}=2}^{n-2}\frac{\bar{\eta}_{\tilde{j}}}{f}\frac{\partial f}{\partial\bar{\eta}_{\tilde{j}}}\Big(\sum_{i=1,i\neq1}^{n}\sum_{j>i}{z_{ij}^{2}}{\cal I}_{221}(z_{i},z_{j},\bar{z}_{1})\frac{\bar{z}_{\tilde{j},n-1}}{\bar{z}_{n-1,1}\bar{z}_{\tilde{j}1}}\\
+&\sum_{i=1,i\neq\tilde{j}}^{n}\sum_{j>i,j\neq\tilde{j}}{z_{ij}^{2}}{\cal I}_{221}(z_{i},z_{j},\bar{z}_{\tilde{j}})\frac{\bar{z}_{1,n}}{\bar{z}_{n,\tilde{j}}\bar{z}_{1\tilde{j}}}
+\sum_{i=1,i\neq\tilde{j}}^{n}\sum_{j=\tilde{j}>i}{z_{ij}^{2}}{\cal I}_{221}(z_{i},z_{\tilde{j}},\bar{z}_{\tilde{j}})\frac{\bar{z}_{1n}}{\bar{z}_{n\tilde{j}}\bar{z}_{1\tilde{j}}}\\
+&\sum_{i=1,i\neq n-1}^{n}\sum_{j>i,j\neq n-1}{z_{ij}^{2}}{\cal I}_{221}(z_{i},z_{j},\bar{z}_{n-1})\frac{\bar{z}_{n}-\bar{z}_{1}}{\bar{z}_{1,n-1}\bar{z}_{n,n-1}}\\+&\sum_{i=1,i\neq n-1}^{n}\sum_{j=n-1>i}{z_{i,n-1}^{2}}{\cal I}_{221}(z_{i},z_{n-1},\bar{z}_{n-1})\frac{\bar{z}_{n}-\bar{z}_{1}}{\bar{z}_{1,n-1}\bar{z}_{n,n-1}}\\+&\sum_{i=1}^{n}\sum_{j>i,j\neq n}{z_{ij}^{2}}{\cal I}_{221}(z_{i},z_{j},\bar{z}_{n})\frac{\bar{z}_{n-1,\tilde{j}}}{\bar{z}_{\tilde{j}n}\bar{z}_{n-1,n}}
+\sum_{i=1}^{n}\sum_{j=n>i}{z_{in}^{2}}{\cal I}_{221}(z_{i},z_{n},\bar{z}_{n})\frac{\bar{z}_{n-1,\tilde{j}}}{\bar{z}_{\tilde{j}n}\bar{z}_{n-1,n}}\\
+&\sum_{j>i=\tilde{j}}{z_{ij}^{2}}{\cal I}_{221}(z_{\tilde{j}},z_{j},\bar{z}_{j})\frac{\bar{z}_{1n}}{\bar{z}_{n\tilde{j}}\bar{z}_{1\tilde{j}}}
+\sum_{j>i}{z_{ij}^{2}}{\cal I}_{221}(z_{1},z_{j},\bar{z}_{1})\frac{\bar{z}_{\tilde{j},n-1}}{\bar{z}_{n-1,1}\bar{z}_{\tilde{j}1}}\\
+&{z_{n-1,n}^{2}}{\cal I}_{221}(z_{n-1},z_{n},\bar{z}_{n-1})\frac{\bar{z}_{n1}}{\bar{z}_{1,n-1}\bar{z}_{n,n-1}}\Big).
\ea
\ee
where ${\cal I}_{221}$ is defined by eq.(\ref{I1111}).

\section{Useful integrals}\label{sec:integral}
It is convenient to define the following notation
\be
\ba
&\mathcal{I}_{a_{1}, \cdots, a_{m}, b_{1}, \cdots, b_{n}}\left(z_{i_{1}}, \cdots, z_{i_{m}}, \bar{z}_{j_{1}}, \cdots, \bar{z}_{j_{n}}\right)\\:=&\int \frac{d^{2} z}{\left(z-z_{i_{1}}\right)^{a_{1}} \cdots\left(z-z_{i_{m}}\right)^{a_{m}}\left(\bar{z}-\bar{z}_{j_{1}}\right)^{b_{1}} \cdots\left(\bar{z}-\bar{z}_{j_{n}}\right)^{b_{n}}}
\ea
\ee
For example, we can write
$$
\begin{aligned}
\mathcal{I}_{2222}\left(z_{1}, z_{2}, \bar{z}_{1}, \bar{z}_{2}\right) &=\int \frac{d^{2} z}{\left|z-z_{1}\right|^{4}\left|z-z_{2}\right|^{4}} \\
\mathcal{I}_{2222}\left(z_{1}, z_{2}, \bar{z}_{3}, \bar{z}_{4}\right) &=\int \frac{d^{2} z}{\left(z-z_{1}\right)^{2}\left(z-z_{2}\right)^{2}\left(\bar{z}-\bar{z}_{3}\right)^{2}\left(\bar{z}-\bar{z}_{4}\right)^{2}} \\
\mathcal{I}_{221111}\left(z_{1}, z_{2}, \bar{z}_{1}, \bar{z}_{2}, \bar{z}_{3}, \bar{z}_{4}\right) &=\int \frac{d^{2} z}{\left(z-z_{1}\right)^{2}\left(z-z_{2}\right)^{2}\left(\bar{z}-\bar{z}_{1}\right)\left(\bar{z}-\bar{z}_{2}\right)\left(\bar{z}-\bar{z}_{3}\right)\left(\bar{z}-\bar{z}_{4}\right)} \\
\mathcal{I}_{11111111}\left(z_{1}, z_{2}, z_{3}, z_{4}, \bar{z}_{1}, \bar{z}_{2}, \bar{z}_{3}, \bar{z}_{4}\right) &=\int \frac{d^{2} z}{\prod_{i=1}^{4}\left(z-z_{i}\right) \prod_{j=1}^{4}\left(\bar{z}-\bar{z}_{j}\right)}.
\end{aligned}
$$

The first particular integral is
$$
{\cal I}_{i, j}\left(z_i, \bar{z}_j\right) \equiv \int \frac{d^{2} z_{4}}{z_{4 i} \bar{z}_{4 j}}
$$
For definiteness, one can compute ${\cal I}_{12}\left(z_1, \bar{z}_2\right)$
\be
\ba
{\cal I}_{1,2}\left(z_1, \bar{z}_2\right)&=\int \frac{d^{2} z_{4} \bar{z}_{41} z_{42}}{\left|z_{41}\right|^{2}\left|z_{42}\right|^{2}}=\int_{0}^{1} d u \int \frac{d^{2} z_{4} \bar{z}_{41} z_{42}}{\left[u\left|z_{41}\right|^{2}+(1-u)\left|z_{42}\right|^{2}\right]^{2}}\\&=\int_{0}^{1} d u \int \frac{d^{2} z_{4}^{\prime}\left(\bar{z}_{4}^{\prime}-(1-u) \bar{z}_{12}\right)\left(z_{4}^{\prime}+u z_{12}\right)}{\left[z_{4}^{\prime 2}+u(1-u)\left|z_{12}\right|^{2}\right]^{2}}
\ea
\ee
Changing the dimension to $d$ one can find
\be\label{I12}
\begin{aligned}
{\cal I}_{1,2}=& 2 V_{S^{d-1}} \int_{0}^{1} d u \int_{0}^{\infty} d \rho \frac{\rho^{d-1}\left(\rho^{2}-u(1-u)\left|z_{12}\right|^{2}\right)}{\left(\rho^{2}+u(1-u)\left|z_{12}\right|^{2}\right)^{2}}\\=&2 \frac{2 \pi^{d / 2}}{\Gamma\left(\frac{d}{2}\right)} \int_{0}^{1} d u \frac{(d-1)\left[u(1-u)\left|z_{12}\right|^{2}\right]^{\frac{d}{2}-1}}{2} \Gamma\left(\frac{d}{2}\right) \Gamma\left(1-\frac{d}{2}\right) \\
=&2 \pi^{d / 2}\left|z_{12}\right|^{d-2} \frac{\Gamma(d / 2)^{2} \Gamma(1-d / 2)}{\Gamma(d-1)}=-2 \pi\left(\frac{2}{\epsilon}+\ln \left|z_{12}\right|^{2}+\gamma_{E}+\ln \pi+\mathcal{O}(\epsilon)\right)
\end{aligned}\ee
Differentiating this result, one can obtain the following simple integrals
$$
\int \frac{d^{2} z_{4}}{z_{43}^{2} \bar{z}_{41}}=\frac{2 \pi}{z_{13}}, \quad \int \frac{d^{2} z_{4}}{z_{43}^{2} \bar{z}_{41}^{2}}=4 \pi^{2} \delta^{2}\left(z_{13}\right)
$$
where one uses $\bar{\partial} \frac{1}{z}=\partial \frac{1}{\bar z}=2 \pi \delta(z) \delta(z) .$

To define $\mathcal{I}_{2222}$, one can do the Feynman parametrization to integrate
\begin{equation}
\begin{array}{l}
\mathcal{I}_{2222}\left(z_{1}, z_{2}, \bar{z}_{1}, \bar{z}_{2}\right)=\int d z^{2} \frac{1}{\left(\left|z-z_{1}\right|^{2}\left|z-z_{2}\right|^{2}\right)^{2}} \\
=6 \int_{0}^{1} d u \int d z^{2} \frac{u(1-u)}{\left(u\left|z-z_{1}\right|^{2}+(1-u)\left|z-z_{2}\right|^{2}\right)^{4}} \\
=6 \int_{0}^{1} d u \int d \tilde{z}^{2} \frac{u(1-u)}{\left(|\tilde{z}|^{2}+u(1-u)\left|z_{12}\right|^{2}\right)^{4}} \\
=12 V_{S^{2-1}} \int_{0}^{1} d u u(1-u) \int_{0}^{\infty} \frac{\rho^{2-1} d \rho}{\left(\rho^{2}+u(1-u)\left|z_{12}\right|^{2}\right)^{4}}
\end{array}
\end{equation}
To regulate the divergence, we use the dimensional regularization by replacing $2 \mathrm{D}$ to
$d \mathrm{D}$
\begin{equation}
\begin{aligned}
\mathcal{I}_{2222}^{(d)}\left(z_{1}, z_{2}, \bar{z}_{1}, \bar{z}_{2}\right)&=12 V_{S^{d-1}} \int_{0}^{1} d u u(1-u) \int_{0}^{\infty} \frac{\rho^{d-1} d \rho}{\left(\rho^{2}+u(1-u)\left|z_{12}\right|^{2}\right)^{4}}\\
&=V_{S^{d-1}} \Gamma\left(2-\frac{d}{2}\right) \Gamma\left(\frac{d}{2}\right)\left|z_{12}\right|^{d-8} \int_{0}^{1} d u \frac{(d-6)(d-4)}{4}(u(1-u))^{d / 2-3}\\
&=2 \pi^{\frac{d}{2}} \Gamma\left(4-\frac{d}{2}\right) B\left(\frac{d}{2}-2, \frac{d}{2}-2 \mid\right)\left|z_{12}\right|^{d-8}\\
&\xrightarrow{{d=2+\tilde{\epsilon}}{{}}} \frac{8 \pi}{\left|z_{12}\right|^{6}}\left(\frac{4}{\tilde{\epsilon}}+2 \log \left|z_{12}\right|^{2}+2 \log \pi+2 \gamma-5\right).
\end{aligned}
\end{equation}

We have also used the following integral
\be
\begin{aligned}\label{I1111}
\mathcal{I}_{221}\left(z_{1}, z_{2}, \bar{z}_{1}\right)=\int d^{2} z \frac{1}{\left(z-z_{1}\right)^{2}\left(z-z_{2}\right)^{2}\left(\bar{z}-\bar{z}_{1}\right)}, \\
\mathcal{I}_{221}\left(z_{1}, z_{2}, \bar{z}_{3}\right)=\int d^{2} z \frac{1}{\left(z-z_{1}\right)^{2}\left(z-z_{2}\right)^{2}\left(\bar{z}-\bar{z}_{3}\right)}.
\end{aligned}
\ee
By using $\partial_{z_{1}} \partial_{z_{2}}\left(\frac{1}{z_{12}}\left(\frac{1}{z-z_{1}}-\frac{1}{z-z_{2}}\right)\right)=\frac{1}{\left(z-z_{1}\right)^{2}\left(z-z_{2}\right)^{2}},$ we find
\be\begin{aligned} & \mathcal{I}_{221}\left(z_{1}, z_{2}, \bar{z}_{1}\right) \\=& \int \frac{d^{2} z}{\left(z-z_{1}\right)^{2}\left(z-z_{2}\right)^{2}\left(\bar{z}-\bar{z}_{1}\right)} \\=& \frac{1}{z_{12}^{2}} \int \frac{d^{2} z}{\left(\bar{z}-\bar{z}_{1}\right)}\left(\frac{1}{\left(z-z_{1}\right)^{2}}+\frac{1}{\left(z-z_{2}\right)^{2}}-\frac{2}{\left(z-z_{1}\right)\left(z-z_{2}\right)}\right) \\=& \frac{1}{z_{12}^{2}} \partial_{z_{2}} \mathcal{I}_{2,1}\left(z_{2}, \bar{z}_{1}\right)-\frac{2}{z_{12}^{3}}\left(\mathcal{I}_{1,1}\left(z_{1}, \bar{z}_{1}\right)-\mathcal{I}_{2,1}\left(z_{2}, \bar{z}_{1}\right)\right). \end{aligned}\ee
Moreover, we have
$$
\begin{aligned}
\mathcal{I}_{221}\left(z_{1}, z_{2}, \bar{z}_{3}\right) &=\int d^{2} z \frac{1}{\left(\bar{z}-\bar{z}_{3}\right)} \partial_{z_{1}} \partial_{z_{2}}\left(\frac{1}{z_{12}}\left(\frac{1}{z-z_{1}}-\frac{1}{z-z_{2}}\right)\right) \\
&=\partial_{z_{1}} \partial_{z_{2}}\left(\frac{1}{z_{12}} \int d^{2} z \frac{1}{\left(\bar{z}-\bar{z}_{3}\right)}\left(\frac{1}{z-z_{1}}-\frac{1}{z-z_{2}}\right)\right) \\
&=\partial_{z_{1}} \partial_{z_{2}}\left(\frac{1}{z_{12}}\left(\mathcal{I}_{1,3}\left(z_{1}, \bar{z}_{3}\right)-\mathcal{I}_{2,3}\left(z_{2}, \bar{z}_{3}\right)\right)\right).
\end{aligned}
$$

One makes use of above notation to reproduce the $T\bar{T}$ and $J\bar{T}$ deformed four-point function given in \cite{He:2019vzf}. For an example, the{ $J\bar{T}-$ first order deformed four-point correlation function eq.(65) in \cite{He:2019vzf} can be rephrased using $\mathcal{I}_{i, j}$ given in eq.(\ref{JT-npoint}) in terms of the following relations
\begin{equation}
\begin{aligned}\label{JTconvention}
\mathcal{I}_{122}\left(z_{1}, \bar{z}_{3}, \bar{z}_{4}\right) &=\int d^{2} z \frac{1}{\left(z-z_{1}\right)\left(\bar{z}-\bar{z}_{3}\right)^{2}\left(\bar{z}-\bar{z}_{4}\right)^{2}} \\
&=\partial_{\bar{z}_{4}} \partial_{\bar{z}_{3}}\left(\frac{1}{\bar{z}_{34}}\left(\mathcal{I}_{1,3}\left(z_{1}, \bar{z}_{3}\right)-\mathcal{I}_{1,4}\left(z_{1}, \bar{z}_{4}\right)\right)\right),\\
\mathcal{I}_{122}\left(z_{1}, \bar{z}_{1}, \bar{z}_{3}\right) &=\int d^{2} z \frac{1}{\left(z-z_{1}\right)\left(\bar{z}-\bar{z}_{1}\right)^{2}\left(\bar{z}-\bar{z}_{3}\right)^{2}} \\
&=\frac{1}{\bar{z}_{13}^{2}}\left(-\frac{2}{\bar{z}_{13}} \mathcal{I}_{1,1}\left(z_{1}, \bar{z}_{1}\right)+\partial_{\bar{z}_{3}} \mathcal{I}_{1,3}\left(z_{1}, \bar{z}_{3}\right)+\frac{2}{\bar{z}_{13}} \mathcal{I}_{1,3}\left(z_{1}, \bar{z}_{3}\right)\right),\\
\mathcal{I}_{11111}\left(z_{1}, \bar{z}_{1}, \bar{z}_{2}, \bar{z}_{3}, \bar{z}_{4}\right)
=& \int d^{2} z \frac{1}{\left(z-z_{1}\right)\left(\bar{z}-\bar{z}_{1}\right)\left(\bar{z}-\bar{z}_{2}\right)\left(\bar{z}-\bar{z}_{3}\right)\left(\bar{z}-\bar{z}_{4}\right)} \\
=&\Big(\frac{1}{\bar{z}_{12} \bar{z}_{13} \bar{z}_{14}} \mathcal{I}_{1,1}\left(z_{1}, \bar{z}_{1}\right)-\frac{1}{\bar{z}_{12} \bar{z}_{23} \bar{z}_{24}} \mathcal{I}_{1,2}\left(z_{1}, \bar{z}_{2}\right)\\+&\frac{1}{\bar{z}_{34} \bar{z}_{13} \bar{z}_{23}} \mathcal{I}_{1,3}\left(z_{1}, \bar{z}_{3}\right)-\frac{1}{\bar{z}_{34} \bar{z}_{14} \bar{z}_{24}} \mathcal{I}_{1,4}\left(z_{1}, \bar{z}_{4}\right)\Big),
\end{aligned}
\end{equation} where $\mathcal{I}_{122}$ and $\mathcal{I}_{11111}$ presented in the \cite{He:2019vzf} are expressed by $\mathcal{I}_{i,j}$ given here.}

\bibliographystyle{JHEP}
\providecommand{\href}[2]{#2}\begingroup\raggedright\endgroup

\end{document}